\documentclass[journal=aelccp,manuscript=letter]{achemso}

\usepackage{amsmath}
\usepackage[version=3]{mhchem} 
\usepackage{enumitem}

\newcommand*\kB{k_{\text{B}}}
\newcommand*\kin{_{\text{kin}}}
\newcommand*\thermo{_{\text{thermo}}}
\newcommand*\kr{k_{\text r}}
\newcommand*\ko{k_{\text o}}
\newcommand*\ar{a_{\text r}}
\newcommand*\ao{a_{\text o}}

\author{Rachel C. Kurchin}
    \email{rkurchin@cmu.edu}
    \affiliation{Carnegie Mellon University}
\author{Dhairya Gandhi}
    \affiliation{Julia Computing}
\author{Venkatasubramanian Viswanathan}
    \email{venkvis@cmu.edu}
    \affiliation{Carnegie Mellon University}

\title[shorttitle]{Nonequilibrium Electrochemical Phase Maps: Beyond Butler-Volmer Kinetics}

\begin{document}

\begin{tocentry}
\centering
\includegraphics[width=8cm]{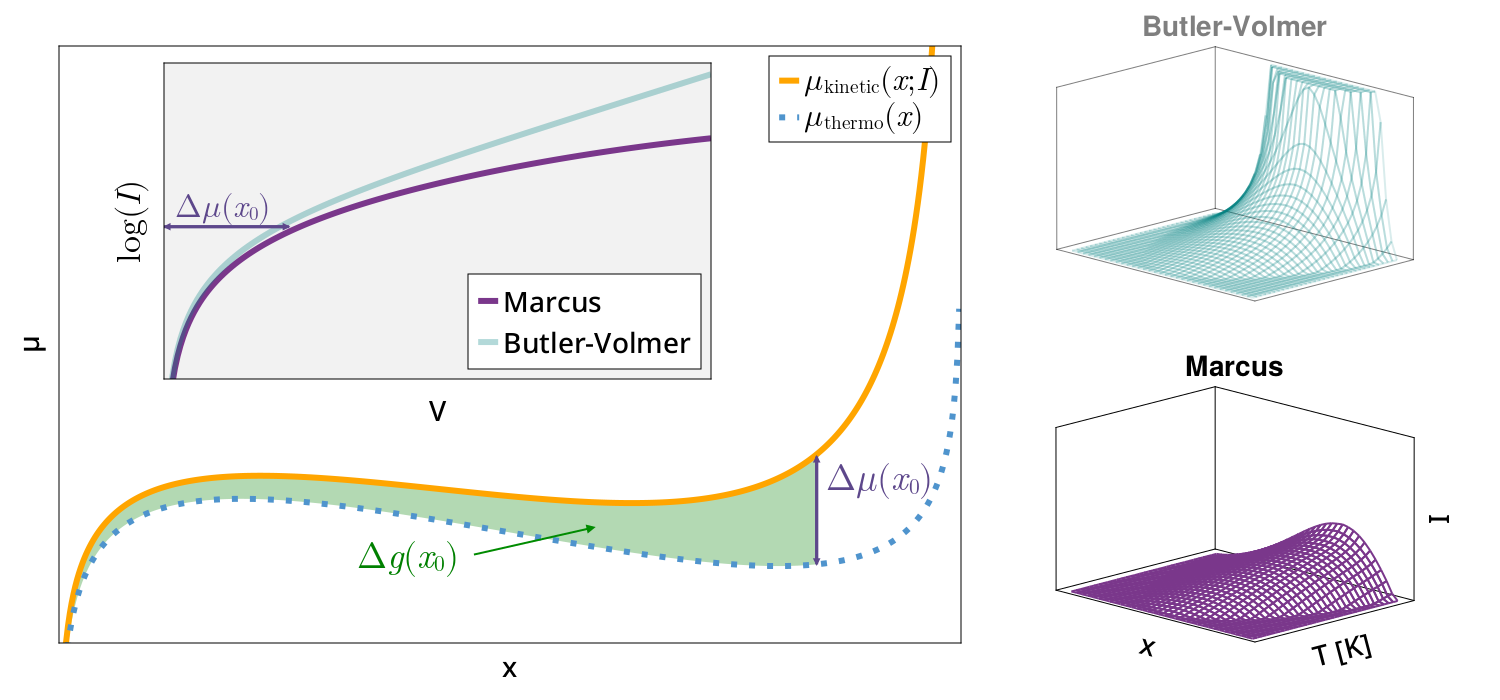}
\end{tocentry}

\begin{abstract}
Electrochemical kinetics at electrode-electrolyte interfaces are crucial to understand high-rate behavior of energy storage devices. Phase transformation of electrodes is typically treated under equilibrium thermodynamic conditions, while realistic operation is at finite rates. Analyzing phase transformations under nonequilibrium conditions requires integrating nonlinear electrochemical kinetic models with thermodynamic models. This had only previously been demonstrated for Butler-Volmer kinetics, where it can be done analytically. In this work, we develop a kinetic modeling package in the Julia language capable of efficient numerical inversion of rate relationships for general kinetic models using automatic differentiation. We demonstrate building nonequilibrium phase maps, including for models such as Marcus-Hush-Chidsey that require computation of an integral, and also discuss the impact of a variety of assumptions and model parameters (such as temperature, reorganization energy, activity, and ideal solution interaction energy), particularly on high-rate phase behavior. Even for a fixed set of parameters, the magnitude of the critical current can vary by in excess of a factor of two amongst kinetic models.
\end{abstract}

Energy storage via batteries plays a vital role in the electrification transformation that is crucial to address climate change. Fast charging capabilities are important to meet consumer demand in passenger electric vehicles~\cite{Crabtree2019}, and fast discharge will be pivotal to enable emerging applications such as electric aviation~\cite{eVTOL,batt_flight}. Modeling the behavior of electrochemical systems at these high rates necessitates going beyond Butler-Volmer (BV) kinetics, as BV is a first-order approximation to the activation energy of reaction valid only at small overpotentials. The larger overpotentials required for high-rate applications requires adoption of higher-order (and more physically interpretable~\cite{henstridge2012marcus}) models such as Marcus theory~\cite{Marcus}, Marcus-Hush-Chidsey (MHC) models~\cite{Chidsey1991}, or our recently-introduced modification, Marcus-Hush-Chidsey-Kurchin-Viswanathan (MHC-KV) that incorporates electrode density of states (DOS) explicitly.~\cite{MHCKV}. In addition, even at lower rates, a variety of systems have been shown to deviate from BV kinetics~\cite{bai2014charge, martinez2020kinetic,laborda2013asymmetric}.

Beyond deviation from BV, high rates can also impact phase behavior~\cite{delacourt2005existence, niu2014situ, hess2015combined, li2018review, li2018orientation}, causing deviation from thermodynamic phase diagrams. Previous work modeled this for the case of BV kinetics of intercalation into \ce{LiFePO4} (LFP) nanoparticles~\cite{Bai2011}. Here, we show for the first time the capability to build nonequilibrium phase maps for any kinetic model with any underlying thermodynamic parameters. To do so, we introduce ElectrochemicalKinetics.jl~\cite{EK}, a Julia language package that provides a common interface for a variety of rate models, including BV, Marcus, MHC, Zeng et al.'s asymptotic approximation to MHC~\cite{Zeng2014}, and our DOS-dependent model. Implementation of additional rate laws (such as asymmetric Marcus kinetics~\cite{Laborda2012, Zeng2015}) is straightforward. Crucially, ElectrochemicalKinetics.jl can not only evaluate them in the ``forward'' direction (computing a current for a given overpotential, $I(\eta)$), but also makes use of automatic differentiation (AD) to efficiently ``invert'' them, computing the overpotential required to drive a given current under a particular model, $\eta(I)$. This ``inverse'' function is needed for building phase maps in the general case, where, unlike BV kinetics with symmetric electron transfer, there is no closed-form inverse.

Continuing with the example of Li intercalation into LFP, we consider the reaction:
\begin{equation}
    x\ce{Li+}+x\ce{e- + FePO4 <=>C[\kr][\ko] Li}_x\ce{FePO4},
\end{equation}

\noindent where $x$ represents a concentration of Li in iron phosphate (and $1-x$ the concentration of ``vacant'' sites). As has been done previously,\cite{Bai2011} we model the thermodynamics via a regular solution model, with an interaction parameter $\Omega$ describing the energy of interaction between an intercalated Li and a vacant site. The molar Gibbs free energy of mixing is then given by
\begin{equation}
    g\thermo(x) = h(x) + Ts(x) = \Omega x(1-x) + \kB T\big(x\ln x + (1-x)\ln(1-x)\big).
    \label{g_thermo}
\end{equation}

Then, by definition, the chemical potential is
\begin{equation}
    \mu\thermo(x) \equiv \frac{\partial g}{\partial x} = \Omega(1-2x)+\kB T\ln\left(\frac{x}{1-x}\right).
    \label{mu_thermo}
\end{equation}

\begin{figure}
    \centering
    \includegraphics[width=\textwidth]{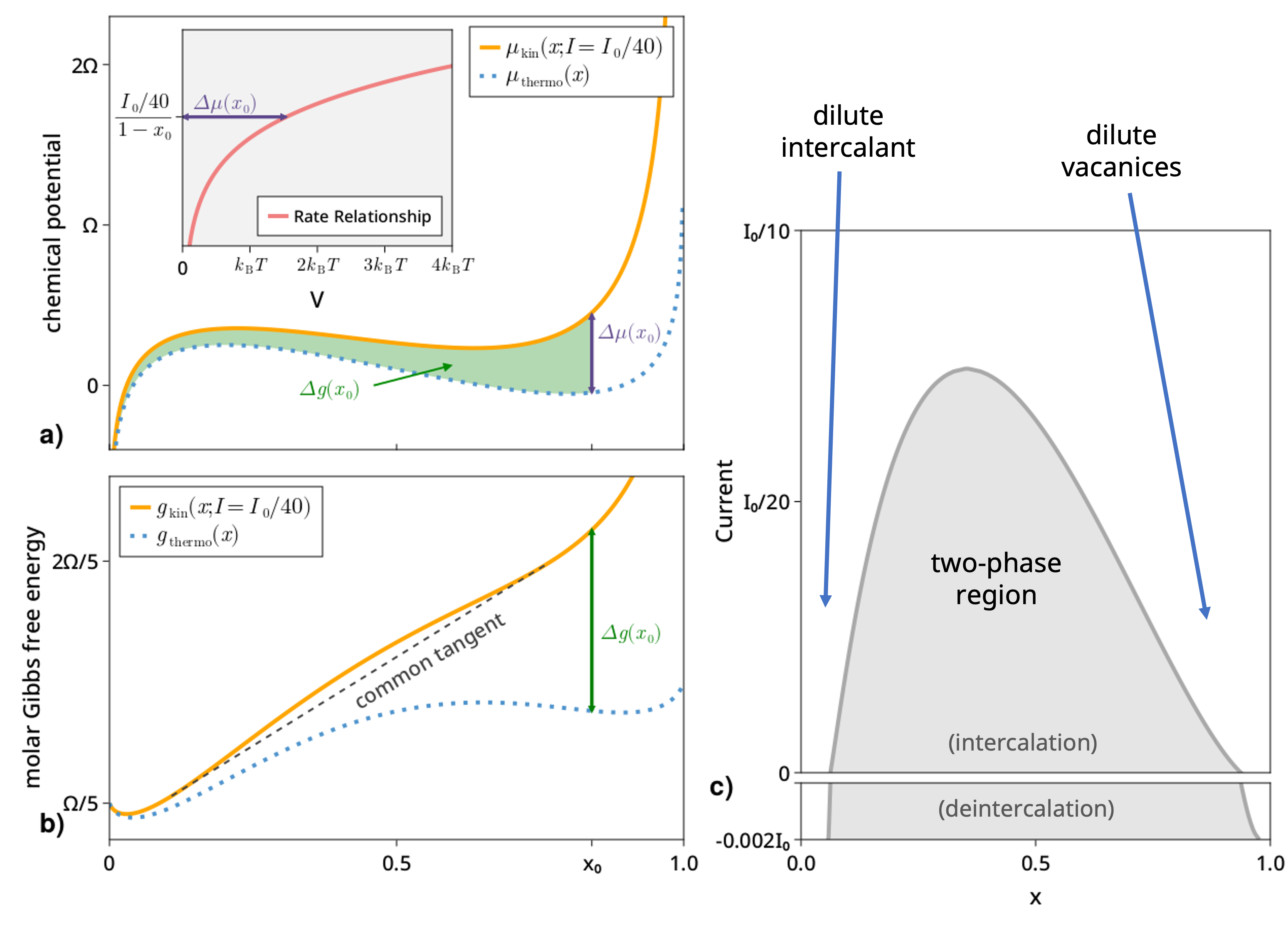}
    \caption{Nonequilibrium phase map construction. a) Kinetic chemical potential is computed for a given rate model (inset) at a given current, here $I_0/40$, where $I_0$ is the maximum current possible for the Marcus model used. b) Chemical potential is integrated to compute the associated molar Gibbs free energy. Phase boundaries at the imposed current are computed using the common tangent condition. c) This process is repeated at different currents to build up the phase map. Schematics of phase behavior in the three regions are superimposed. Stripping is plotted as negative current. For these plots, the interaction parameter $\Omega\simeq3\kB T$ and reorganization energy $\lambda\simeq10\kB T$.}
    \label{fig1}
\end{figure}

To incorporate the effects of a steady-state intercalation current $I$, we modify Equation~\ref{mu_thermo} by adding a term to account for the overpotential needed to drive that current:
\begin{equation}
    \mu\kin(x) = \mu\thermo(x) + \eta\left(I\right)
    \label{mu_kin}
\end{equation}
The reaction rate (and thereby the current) has two parts: the intrinsic reaction rate of the process and available sites for the reaction~\cite{hansen2014unifying}. This necessitates a model for the activities of the intercalated Li atoms and the vacant sites; denote these as $\ar(x)$ and $\ao(x)$ for now (we discuss possible choices for these functions and their implications below). The current is then given by
\begin{equation}
    I = \ao(x) \ko(\eta) - \ar(x) \kr(\eta),
\end{equation}
where we have adopted the convention of oxidative current being positive, and the functions $\ko(\eta)$ and $\kr(\eta)$ represent the overpotential dependence of the rate constants in the oxidative and reductive directions under otherwise standard conditions (all nondimensional concentrations unity).
In our analysis, we will assume constant concentration of Li ions. 

Bai et al.~\cite{Bai2011} used $\ao(x) = \ar(x) = 1-x$, based on an excluded volume argument. Assuming these activities and a Butler-Volmer rate law for $\ko$ and $\kr$ leads Equation~\ref{mu_kin} to become equivalent to Equation 9 in Ref.~\citenum{Bai2011}, but there the $\sinh^{-1}$ term comes from an analytical inverse of BV kinetics with symmetric electron transfer, whereas here we make use of the \verb|overpotential| function in ElectrochemicalKinetics to invert any rate model with any model for activity (see below and SI for more details).

Since it still must be true that $\mu(x)=\frac{\partial g}{\partial x}$, we then also have that
\begin{equation}
    g\kin(x) = g\thermo(x) + \int_0^x\hspace{-2mm}\mu\kin(x')dx'.
\end{equation}
With this expression, we can use the usual common tangent construction to find the phase boundaries as a function of the imposed current. (Note that our analysis here presumes that intercalation is kinetically limited as opposed to transport-limited. In future work we plan to couple the software to a transport model.)

Figure~\ref{fig1} shows plots of $\mu\thermo(x)$, $\mu\kin(x)$, $g\thermo(x)$, and $g\kin(x)$ for $1-x$ activities and a Marcus model (Tafel plot in inset), as well as the constructed nonequilibrium phase map, with intercalation currents plotted as positive and deintercalation as negative. $I_0$ is the maximum value the Marcus current can take on (just before the start of the inverted region), and the plots of $\mu$ and $g$ are computed at a current of $I_0/40$.

For more details on how these calculations are accomplished, see the Supporting Information. Here, we remark that the speed of the Julia language in general~\cite{bezanson2017julia}, and especially automatic differentiation (AD) support via the Zygote~\cite{zygote} package, are crucial to the ability to construct these phase maps within feasible times on a personal computer. In particular, in order to invert a generic rate model, a numerical solve is necessary (since no analytical inverse exists in general), and it is crucial that this calculation be efficient, since the \verb|overpotential| function is called many times (and is itself optimized over) in determining phase boundaries at a given current in order to build a phase stability map. AD allows this solve to proceed via a gradient-based optimization without the need for finite differencing, which, especially for more complicated rate models, can enable substantially fewer optimizer steps.

\begin{figure}
    \centering
    \includegraphics[width=\textwidth]{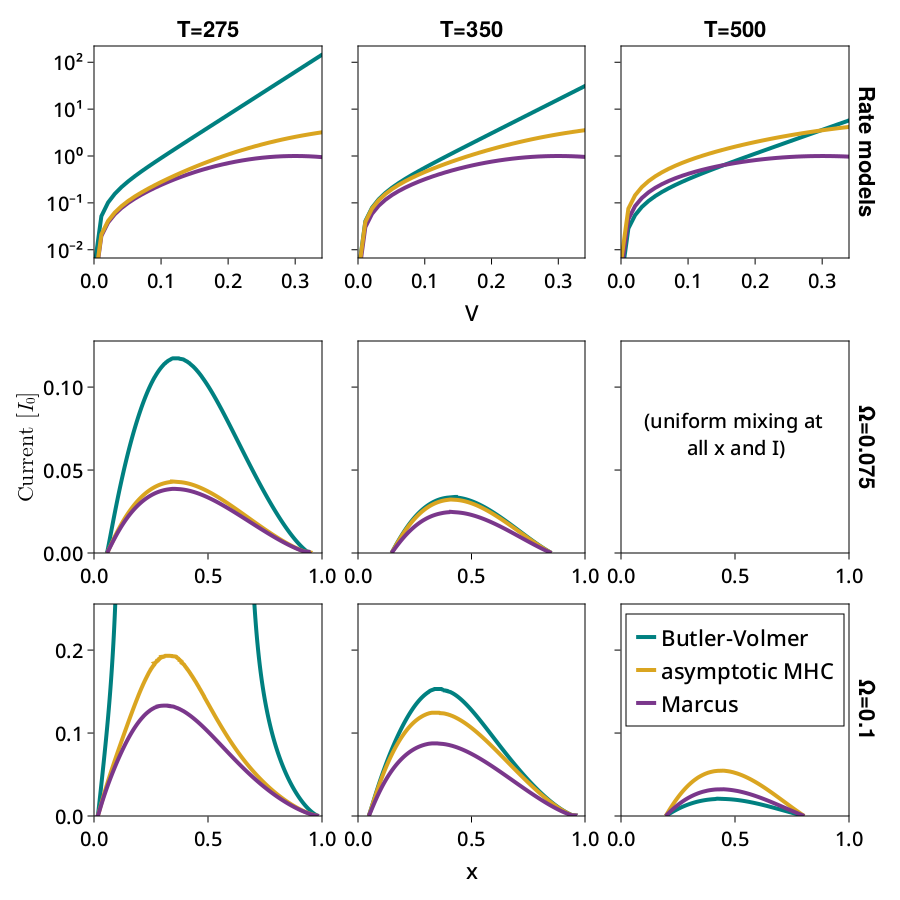}
    \caption{Phase maps for intercalation under three different rate models at three different temperatures (Tafel plots shown in top row) and two different interaction parameters $\Omega$ (second and third rows). For the Marcus and MHC models, $\lambda=0.3$ eV. Currents are nondimensionalized by the prefactor $I_0$ of the Marcus model, and prefactors of all models chosen such that low-overpotential Tafel plots coincide at 300K.}
    \label{fig2}
\end{figure}

For the remainder of this work, we will focus on the case of intercalation, where larger currents suppress phase separation. We now turn to the impact of the thermodynamic parameters (in the case of ideal mixing, the interaction parameter $\Omega$) and kinetic model. We choose three temperatures and two values of the interaction parameter to demonstrate the qualitative range of observed behaviors. We parameterize three models (Butler-Volmer, Marcus, and asymptotic MHC) such that their low-overpotential behavior at room temperature is equivalent, and then fix these parameters. Figure~\ref{fig2} illustrates the variation of the phase maps with model type, temperature, and interaction parameter. The Tafel plots for each model at each temperature are shown in the top row, and phase maps for two values of $\Omega$ in the bottom two. (For an investigation of the impact of asymmetric electron transfer in a BV model, see the Supporing Information.)

Several interesting effects are apparent from this parameter sweep. First, there are substantial qualitative and quantitative differences between the phase maps under different kinetic models: most notably, at larger $\Omega$ and lower $T$ for a Butler-Volmer model, there is no critical current beyond which phase separation stops, but rather there is a two-phase region at some intermediate range of compositions no matter how large the current -- this is directly related to the fact that, unlike more physics-based models, a BV model can mathematically reach any output current with a large enough applied overpotential. 


Second, we note the trends in critical current with these parameters. It increases with increasing interaction parameter $\Omega$, and decreases with temperature, eventually yielding uniform mixing at all compositions, even with no current. These are both in fact thermodynamic effects, as the mixing model in Equation~\ref{g_thermo} can easily be shown to have a critical temperature (above which there is uniform mixing) at $\Omega/2\kB$ (i.e. 435K when $\Omega=0.075$ eV and 580K when $\Omega=0.1$ eV), or, equivalently, a critical $\Omega$ (below which there is uniform mixing) of $2\kB T$ (i.e. 0.043, 0.060, and 0.086 eV for temperatures of 250, 350, and 500K, respectively). We also note that the trend of critical current with temperature is quite different for the three different models –- critical current decreases substantially faster for BV models, such that the rank ordering between the three changes as $T$ increases.

Figure~\ref{fig2} is, of course, simply a series of slices out of what could be thought of as a three-dimensional phase map, with critical current as a function of both $x$ and $T$. Visualizations of these 3D phase maps can be found in the Supporting Information.

\begin{figure}
    \centering
    \includegraphics[width=\textwidth]{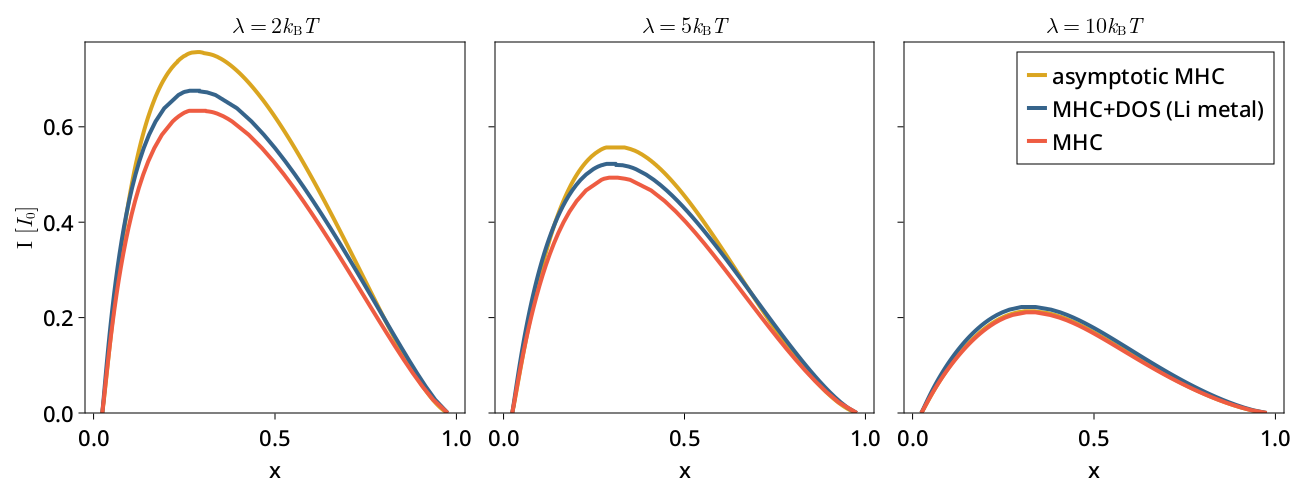}
    \caption{phase maps for two integral-based MHC-based models at a variety of reorganization energies $\lambda$. Asymptotic approximation to MHC also shown for comparison. Currents nondimensionalized as in Figure~\ref{fig2}, by the prefactor $I_0$ of the equivalent Marcus model.}
    \label{fig3}
\end{figure}

So far, we have only considered models that do not require evaluation of an integral to compute rate constants. ElectrochemicalKinetics.jl also supports integral-based models, such as the full (i.e. not asymptotically approximated) MHC model, and our recently introduced MHCKV variant involving the electrode DOS.

Phase maps for these models at three values of the reorganization energy $\lambda$ are shown in Figure~\ref{fig3}, along with the asymptotic approximation to MHC for comparison. Again, some interesting trends can be observed. First, we see a lowering of critical current with increasing reorganization energy, which is not surprising, as $\lambda$ proximally sets the Tafel slope, and hence a larger $\lambda$ means a larger overpotential will be necessary to achieve any given current. In addition, we see that the agreement between the full solution of the MHC model and its asymptotic approximation improves with increasing $\lambda$ as well. This approximation being worse when $\lambda$ is comparable to $\kB T$ is unsurprising, given that its construction involved taking the limits $\lambda\ll\kB T$ and $\lambda\gg\kB T$ and constructing an interpolation between these regimes.~\cite{Zeng2014}

The ability to construct these at all is a remarkable testament to the power of the Julia language, as computing a set of phase boundaries at a single value of the current requires: (i) Optimization through the computation of an integral in order to evaluate the \verb|overpotential| function a single time, (ii) Integration of the \verb|overpotential| function to compute $g\kin(x)$, (iii) Optimization through $g\kin$ to identify the pair of $x$ values that satisfy the common tangent condition.
All of this is accomplished in a few minutes on a single processor.

\begin{figure}
    \centering
    \includegraphics[width=0.4\textwidth]{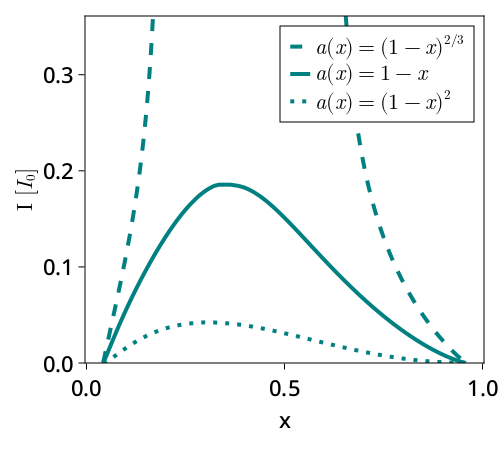}
    \caption{Phase maps for a Butler-Volmer model with three different models for activity: $1-x$, the one assumed by Bai et al.~\cite{Bai2011}, $(1-x)^2$, corresponding to a two-site reaction scheme, and $(1-x)^{2/3}$, a surface reaction.}
    \label{fig4}
\end{figure}

Finally, we return to the matter of the model for the activity of the intercalated vs. nonintercalated sites. This may seem a minor point, but this assumption can have dramatic impacts on the resulting phase map. This is demonstrated in Figure~\ref{fig4}, which shows a phase map for a Butler-Volmer model under three different activity models:
\begin{enumerate}[label=(\roman*)]
    \item $1-x$, the same one used for prior work and prior figures
    \item $(1-x)^{2/3}$, representative of a surface-limited reaction
    \item $(1-x)^2$, representative of a two-site reaction
\end{enumerate}
It is clear from the figure that the nature of the activity has a substantial effect on the phase behavior. The activity model is passed in as a callable function, and so this approach can easily be extended to any function of concentration, such as a Debye-H\"uckel~\cite{huckel1923theorie} or TCPC~\cite{ge2007correlation} model.

To conclude, we have demonstrated the capability to construct nonequilibrium phase maps for any electrochemical rate relationship using the package ElectrochemicalKinetics.jl, which leverages the high performance of the Julia language, as well as its full-featured automatic differentiation capabilities.

The analysis enabled by this functionality compellingly demonstrates the importance of moving beyond simple Butler-Volmer models if accurate models of high-rate behavior are desired. In the future, we plan to integrate the package with battery modeling packages such as PyBAMM~\cite{pybamm} in order to couple it to charge transport (e.g. in a Doyle-Fuller-Newman model~\cite{doyle1993modeling}) and assess the impact on device performance.

All data shown in this work were generated using version 0.2.2 of ElectrochemicalKinetics.jl, and code to generate figures can be found at \verb|https://github.com/BattModels/EK_paper|.

\begin{acknowledgement}


The information, data, or work presented herein was funded in part by the Advanced Research Projects Agency-Energy (ARPA-E), U.S. Department of Energy, under Award Number DE-AR0001211. The views and opinions of authors expressed herein do not necessarily state or reflect those of the United States Government or any agency thereof.

\end{acknowledgement}

\begin{suppinfo}

More details about the software package ElectrochemicalKinetics.jl, examples of 3D phase maps, phase maps for Butler-Volmer models with varying charge transfer coefficients, battery modeling checklist~\cite{Mistry2021}.

\end{suppinfo}

\bibliography{refs}

\end{document}


\section{ElectrochemicalKinetics.jl}
ElectrochemicalKinetics.jl can be found free and open-source on GitHub~\cite{EK}, and is also a registered package in the Julia General registry. In this section, we summarize some of the package's main features, with an emphasis on performance-critical aspects. For more, we refer readers to the \verb|README| file in the repository and a recent JuliaCon talk~\cite{EK_JC}. All data shown in this work were generated using version 0.2.2 of the software.

\subsection{API Overview}
\subsubsection{Computing rates and overpotentials}
Virtually all functionality in the package is, directly or indirectly, built around the \verb|rate_constant| function. It takes two required arguments, the overpotential and a \verb|KineticModel| object (see next subsection), and returns the rate constant at that overpotential within the given model. It takes optional arguments to modify the temperature, and to consider only oxidative or reductive rather than net rates.

The ``inverse'' of \verb|rate_constant| is \verb|overpotential|, which takes in a rate constant/current and a \verb|KineticModel| object and returns the overpotential that would give rise to that current within the given model. It takes an optional argument for temperature, as well as several others that modify some of its internal behavior.

Because most rate models do not have analytical expressions for their inverses, \verb|overpotential|  performs a numerical solve to determine the overpotential. By default, automatic differentiation as implemented in Zygote.jl~\cite{zygote} is used to accomplish this using a gradient-descent optimizer. This functionality can be turned off by passing \verb|autodiff=false| as a keyword argument. The initial guess for the optimizer can also be supplied via the \verb|guess| keyword.

We note also that some models are not one-to-one functions (e.g. a Marcus model that has an inverted region). This can lead to numerical instabilities in some cases, but if reasonable initial guesses are provided, usually does not lead to major issues.

\subsubsection{Model types supported}
\verb|KineticModel| types are divided into \verb|IntegralModel|s and \verb|NonIntegralModel|s. \verb|IntegralModel| types require computation of an integral for a single rate constant computation. These types include:
\begin{itemize}
    \item \verb|MarcusHushChidsey|, a standard Marcus-Hush-Chidsey (MHC)~\cite{Chidsey1991} model parameterized by a prefactor and reorganization energy
    \item \verb|MarcusHushChidseyDOS|, an implementation of our MHC variant variant incorporating electrode density of states (DOS)~\cite{MHCKV}, also includes a prefactor and reorganization energy, and also requires density of states information
\end{itemize}

\verb|NonIntegralModel| types have a closed-form expression for rate constant as a function of overpotential that does \emph{not} require computation of an integral. These types include:
\begin{itemize}
    \item \verb|ButlerVolmer|, a standard Butler-Volmer model parameterized by a prefactor and a charge transfer coefficient
    \item \verb|Marcus|, a Marcus~\cite{Marcus} model parameterized by a prefactor and a reorganization energy
    \item \verb|AsymptoticMarcusHushChidsey|, an implementation of the asymptotic approximation to MHC~\cite{Zeng2014} and parameterized by a prefactor and a reorganization energy
\end{itemize}

The multiple dispatch functionality of Julia means that once we implement \verb|rate_constant| for a given \verb|KineticModel| type, all other functionality such as \verb|overpotential|, as well as features discussed in subsequent sections, will ``just work,'' as the generic implementations only rely on \verb|rate_constant| being dispatched on these types. This also means that ElectrochemicalKinetics.jl is easy to extend with other rate models.

\subsubsection{Phase map construction}
Once the \verb|overpotential| function works efficiently, implementation of Equations 1-5 is relatively straightforward. With these functions in hand, finding phase boundaries for a given value of the current just requires identifying the pair of points $x$ that satisfy the common tangent condition, which can be done with off-the-shelf optimization tools. Construction of a phase map then requires finding these points for a range of current values. Since an optimization problem has to be solved for each current, we speed this up by utilizing the phase boundaries at a given current to inform the starting guess for the optimizer at the next higher value of the current, which cuts down the number of iterations substantially. All this is automated using the \verb|phase_diagram| function.

\begin{figure}
    \centering
    \includegraphics[width=\textwidth]{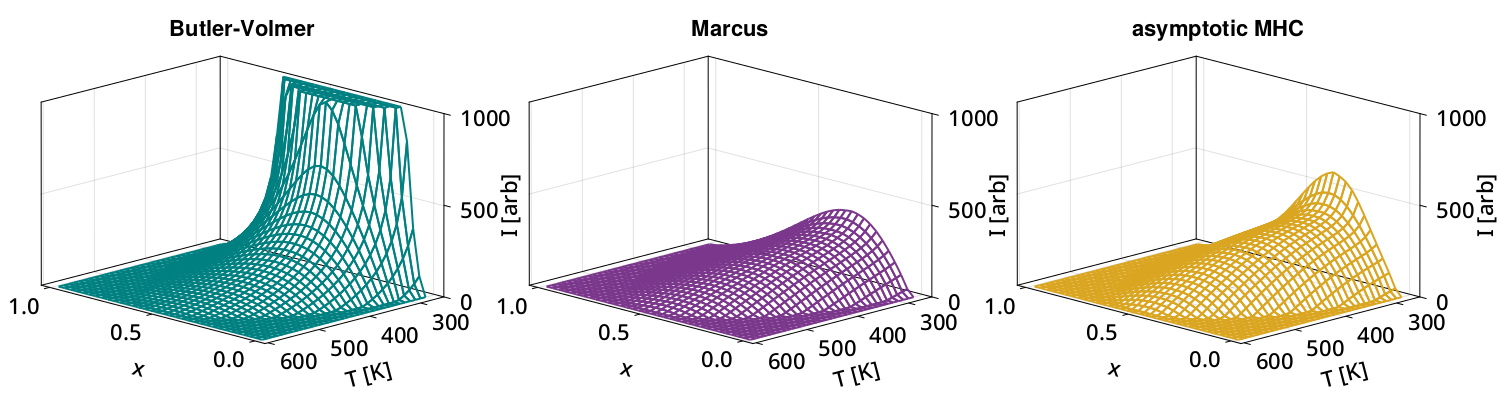}
    \caption{Three-dimensional phase maps for Butler-Volmer, Marcus, and asymptotic MHC models parameterized to have overlapping Tafel plots at low overpotential at 300K.}
    \label{fig:3d}
\end{figure}

Figure~\ref{fig:3d} shows 3D phase maps (as a function of both composition and temperature) for the three non-integral models. The ``floor'' of these plots is the usual thermodynamic phase diagram in $x$ and $T$.

\begin{figure}
    \centering
    \includegraphics[width=0.5\textwidth]{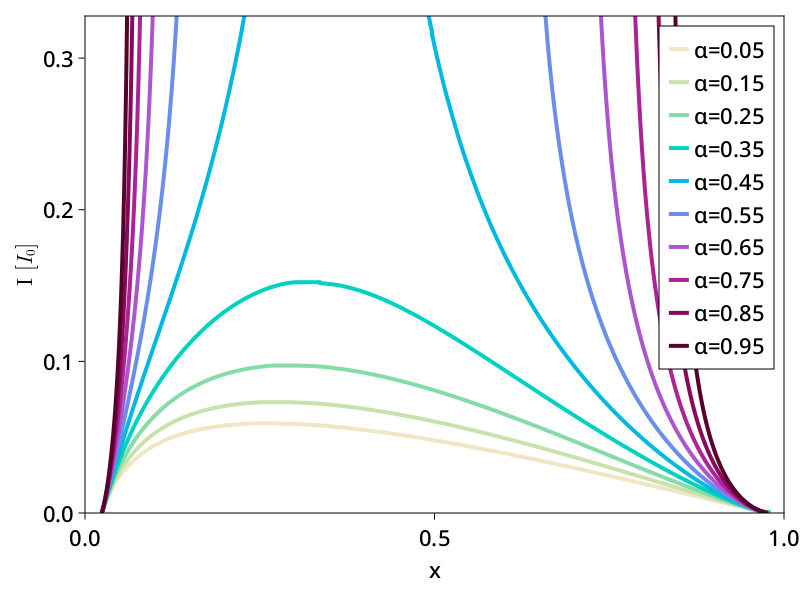}
    \caption{Phase maps at 300K for a Butler-Volmer model with a variety of charge transfer coefficients.}
    \label{fig:etransfer}
\end{figure}

Figure~\ref{fig:etransfer} shows another parameter sweep not included in the main text: namely, the charge transfer coefficient $\alpha$ in a Butler-Volmer model. It has a qualitatively similar effect to temperature, but the actual composition at which the critical current occurs also shifts.

\subsubsection{Other functionality}
The package also has a variety of other functions for plotting/comparing Tafel plots of different rate models, fitting parameters of those models, as well as accounting for the effects of quantum capacitance, important for electrocatalysis at interfaces with 2D materials such as twisted bilayer graphene~\cite{yu2022tunable}.




\newpage
\section{Battery Modeling Checklist}

\begin{table}[!h]
    \begin{tabular}{@{}p{2in}p{4.5in}@{}}
        \toprule
        Manuscript Title: &
        Nonequilibrium Electrochemical Phase Maps: Beyond Butler-Volmer Kinetics 
        \\ \midrule
        Submitting Author${}^{\ast}$: & 
        Rachel Kurchin
        \\ 
    \end{tabular}

    \begin{tabular}{@{}lp{5.5in}c@{}}
        \toprule
        \# & Question & Y/N/NA${}^{\dagger}$ \\ \midrule
        1 & Have you provided all assumptions, theory, governing equations, initial and boundary conditions, material properties, e.g., open circuit potential (with appropriate precision and literature sources), constant states, e.g., temperature, etc.? & 
        Y
        \\
        & \multicolumn{2}{p{6.3in}}{\textbf{Remarks:}}  \\

        \midrule
        2 & If the calculations have a probabilistic component (e.g. Monte Carlo, initial configuration in Molecular Dynamics, etc.), did you provide statistics (mean, standard deviation, confidence interval, etc.) from multiple $\left( \geq 3 \right)$ runs of a representative case? & 
        NA
        \\
        & \multicolumn{2}{p{6.3in}}{\textbf{Remarks:} Calculations do not have a probabilistic component.} \\

        \midrule
        3 & If data-driven calculations are performed (e.g. Machine Learning), did you specify dataset origin, the rationale behind choosing it, what all information does it contain and the specific portion of it being utilized? Have you described the thought process for choosing a specific modeling paradigm?  & 
        NA
        \\
        & \multicolumn{2}{p{6.3in}}{\textbf{Remarks:} No data-driven calculations are performed.} \\

        \midrule
        4 & Have you discussed all sources of potential uncertainty, variability, and errors in the modeling results and their impact on quantitative results and qualitative trends? Have you discussed the sensitivity of modeling (and numerical) inputs such as material properties, time step, domain size, neural network architecture, etc. where they are variable or uncertain? &
        Y
        \\
        & \multicolumn{2}{p{6.3in}}{\textbf{Remarks:}  } \\

        \midrule
        5 & Have you sufficiently discussed new or not widely familiar terminology and descriptors for clarity? Did you use these terms in their appropriate context to avoid misinterpretation? Enumerate these terms in the `Remarks'. & 
        Y
        \\
        & \multicolumn{2}{p{6.3in}}{\textbf{Remarks:} } \\ \bottomrule
    \end{tabular}
    
    {
    \footnotesize
    ${}^{\ast}$ I verify that this form is completed accurately in agreement with all co-authors, to the best of my knowledge.
    }
    
    {
    \footnotesize
    ${}^{\dagger}$ Y $\equiv$ the question is answered completely. Discuss any N or NA response in `Remarks'.
    }
\end{table}

\newpage
\bibliography{refs}